# Exoplanet Characterization and the Search for Life

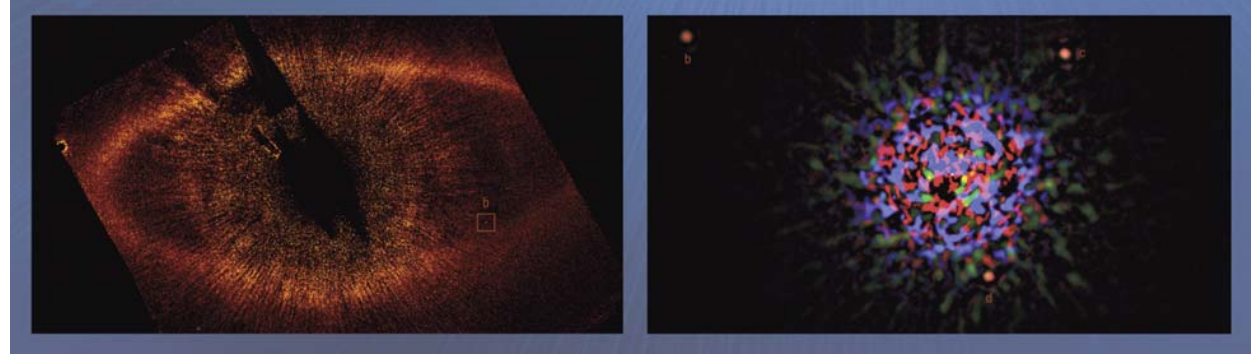

James Kasting, Penn State (lead author)
Alain Leger, France
Amir Vosteen, Netherlands
Angelle Tanner, NExScI/JPL
Ben R. Oppenheimer, AMNH
Carl Grillmair, IPAC
Charles Hanot, Belgium
Christian Marois, Canada
David Montes, Spain
Denis Mourard, France
 Edwin Kite, UC Berkeley
Caltech/CNRS
Garth Illingworth, UCSC
Geoff Marcy, UC Berkeley
Gregory Laughlin, UCO Lick
Henry C. Ferguson, STScI
Jean Surdej, Belgium
John Bally, U. Colorado
John T. Trauger, JPL
Joe Harrington, UCF
Kaspar von Braun, IPAC
K. Balasubramanian, JPL
Marco Spaans, Netherlands
Mark Burchell, UK
Mark S. Marley, ARC
Matthew Kenworthy, U. Ariz.
Ming Zhao, U. Michigan
Nader Haghighipour, U. Hawaii
Nick Woolf, U. Arizona
Olivier Guyon, NAOJ
Pascal Fouque, France
Peter Plavchan, IPAC
Pin Chen, JPL
Rens Waters, Netherlands
Robert A. Brown, STScI
Rolf Danner, NGST
Sara Seager, MIT
S. Sona Hosseini, UC Davis
Stephen Ridgway, NOAO
Stuart Shaklan, JPL
Tsevi Mazeh,Harvard/Israel
William Danchi, GSFC

Wesley A. Traub, JPL (co-author)
Alan Schwartz, Netherlands
Amy Lo, NGST
Athena Coustenis, France
Bertrand Mennesson, France
Charles Beichman, NExScI/JPL
Chris McCarthy, SFSU
Claude Aime, France
David Wilner, CfA
Dimitri Mawet, JPL
Eric Chassefiere, France
Feng Tian, U. Colorado
Gautam Vasisht, JPL
Geoffrey Bryden, JPL
Guillermo Torres, CfA
Hiroshi Shibai, Japan
Jennifer Wiseman, GSFC
John Krist, JPL
Jonti Horner, UK
Jun Nishikawa, Japan
Katia Biazzo Italy
Lisa Kaltenegger, CfA
Margaret Turnbull, GSI
Mark Ealey, NGST
Martin Dominik, UK
M . Muterspaugh, Tennessee State
Motohide Tamura, Japan
Nancy Kiang, GISS
Nigel Mason, UK
Oliver Lay, JPL
Paul Kalas, UC Berkeley
Phil Hinz, U. Arizona
Rachel Akeson, NExScI
Richard Barry, GSFC
R. Vanderbei, Princeton
Ron Allen, STScI
Scott MacPhee, ARC
Stanmir Metchev, SUNY
Stephen Rinehart, GSFC
Theo ten Brummelaar, GSU
Victoria Meadows, U. Wash.
Wing Ip, Taiwan

Aki Roberge, GSFC
Alyn Wooten, NRAO
Andre Brack, France
Benjamin Lane, Draper
Bruno Lopez, France
Charles Cockell, UK
Chris Stark, U. Maryland
D. Angerhausen, Germany
Denis Defrere, Belgium
Douglas Lin, UCSC
Fabien Malbet,
Frances Westall, France
Gene Serabyn, JPL
Glenn White, UK
Heidi B. Hammel, SSI
H. Röttgering, Netherlands
Jian Ge, U. Florida
John Monnier, U. Michigan
Joseph Catanzarite, JPL
Karl  Stapelfeldt, JPL
Kenneth Carpenter, GSFC
Marc Postman, STScI
Marie Levine, JPL
Mark J. Kuchner, GSFC
Matt Mountain, STScI
Michael Shao, JPL
N. Jeremy Kasdin, Princeton
Nicholas Elias, Germany
Olivier Absil, Belgium
Pascal Borde, France
Patrick Lowrance, IPAC
Pierre Kervella, France
Remi Soummer, STScI
Rick Kendrick, Lockheed
Robert Woodruff, Lockheed
Ronald Polidan, NGST
Shri Kulkarni, Caltech
Stella Kafka, Spitzer Sci. Ctr.
Steve Unwin, JPL
Thomas N. Gautier, JPL
Werner Weiss, Austria
Yves Rabbia, France

*Fomalhaut b & HR 8799 b,c,d images:  P. Kalas & C. Marois*     *All signers contributed or requested to be listed.*

# Exoplanet Characterization and the Search for Life

*Abstract*: Over 300 extrasolar planets (exoplanets) have been detected orbiting nearby stars. We now hope to conduct a census of all planets around nearby stars and to characterize their atmospheres and surfaces with spectroscopy. Rocky planets within their star's habitable zones have the highest priority, as these have the potential to harbor life. Our science goal is to find and characterize all nearby exoplanets; this requires that we measure the mass, orbit, and spectroscopic signature of each one at visible and infrared wavelengths. The techniques for doing this are at hand today. Within the decade we could answer long-standing questions about the evolution and nature of other planetary systems, and we could search for clues as to whether life exists elsewhere in our galactic neighborhood.

## *Introduction*

A few times in human history, astronomers have made discoveries that changed people's view of the universe and of themselves. The most renowned of these was Copernicus' suggestion, and Galileo's subsequent proof, that the Earth orbited the Sun, rather than vice versa. This list should also include the discovery that stars are other Suns, that some nebulae are galaxies like our own, and that the universe began with a Big Bang some 13 billion years ago. We now stand at the brink of answering two other paradigm-changing questions: Do other planets like Earth exist, and do any of them harbor life? The tools for answering these questions either exist already or can be developed within the next 10-20 years.

We address here the scientific goal of determining the nature of nearby planetary systems, including giant planets, terrestrial planets, and zodiacal dust, with the expectation that this knowledge will allow us to learn about the evolution of these systems, and will ultimately allow us to discover and examine terrestrial planets. To guide our thinking, and to provide calibration points, we focus on a hypothetical Solar-System-twin at a distance of 10 pc. For such a planet we wish to measure its orbit, mass, visible spectrum, and infrared spectrum, as well as the variation of these quantities with time. From these observations we can potentially derive planetary properties, including effective temperature, radius, mass, density, albedo, atmospheric mass, greenhouse gases, lapse rate, surface gravity, and surface reflectance.

For a hypothetical Earth-twin, we could then estimate its likelihood of habitability, and search for signs of life. Habitability factors include surface temperature, presence of liquid water on the surface, atmospheric pressure, likelihood of plate tectonics, likelihood of atmospheric retention, clouds (cirrus or cumulus), and surface type (rock, ice, sand, water, vegetated). Further factors include the time variability of its observed characteristics, including the length of day, surface morphology (continents, oceans, ice), large-scale weather patterns, obliquity, and seasons. Evidence for life would include the presence of various biogenic trace gases, especially $O_2$ and its photochemical byproduct, $O_3$. Methane is also a potential bioindicator on early-Earth type exoplanets.

## *Search and Characterization Context*

14 years of observations by the radial velocity (RV) technique, as well as gravitational microlensing, transits, and most recently direct imaging[1,2] have revealed over 300 exoplanets. Transiting planets are particularly interesting because some of them can be studied spectroscopically at visible[3] and infrared[4] wavelengths. But transit spectroscopy is likely to be difficult or impossible for a terrestrial planet in its habitable zone (HZ), except possibly for M dwarfs, where the HZ is close to the star. Let us define what we mean by "habitable zone." Many planets and moons could conceivably be habitable, especially as the definition of life itself is not entirely clear. In the Solar System, we can cast our net broadly and include bodies like Mars or Europa that may have subsurface water, as well as bodies like Titan that have complex organic compounds (but no liquid water). These objects can eventually be explored in situ and in great detail. For exoplanets, the possibilities are more limited. Detection of life on exoplanets must be done remotely by observing its effect on the planet's atmosphere or surface. Under these circumstances it is prudent to restrict our search to life as we know it, that is, to organisms that depend on the availability of liquid water. Furthermore, in order for organisms to significantly modify a planet's atmosphere or surface, that water needs to be available at the planet's surface. The planet thus needs to be within the conventional circumstellar HZ, which, for a Sun-like star, extends from roughly 0.8-1.8 AU.[5]

Although some measurements can be made from the ground, most of the science outlined here can only be accomplished by future spaced-based observational techniques. For context, we mention the leading contenders for these techniques here, not as an endorsement, but rather to illustrate that our science aspirations are grounded in extensive engineering analyses and laboratory demonstrations of instruments capable of detecting and characterizing a putative Earth-twin around each of at least the 100 nearest stars. Our reference instrumentation for detecting exoplanets and measuring masses and orbits is the (descoped) Space Interferometer Mission (SIM Lite) in combination with future ground based radial velocity searches that have already detected several planets in the 5 to 10 Earth-mass range [24]. Our reference instrument for measuring visible and near-infrared spectra is the Terrestrial Planet Finder Coronagraph or Occulter (TPF-C/O). Our reference instrument for measuring thermal infrared spectra is the Terrestrial Planet Finder Interferometer (TPF-I) or, in Europe, the Darwin Mission.

*Exoplanet detection, masses, and orbits from astrometry, radial velocity and direct imaging*
A first step would be to identify nearby Earth-twin exoplanets and to measure their masses and orbits. This knowledge is essential to interpreting all of the quantities that can be measured by photometry and spectroscopy. Absent such knowledge, models and estimates can be substituted; however measurements of masses and orbits are, without question, critical to the accurate interpretation of spectroscopic observations.

An estimate of a planet's mass is critical to the issue of habitability. Planets much smaller than 1 $M_\oplus$ may lose heavy gases to space, in addition to H and He. Mars (at ~0.1 $M_\oplus$) is a good example[6]. Small planets also lose their internal heat faster, thereby removing the energy source required to drive plate tectonics. Recycling of carbonate rocks to refresh atmospheric $CO_2$ is thought to be a key factor in stabilizing Earth's long-term climate[5]. Once again, Mars appears to be below the critical mass threshold needed to do this effectively[7]. At the other end of the mass scale, planets larger than about 10 $M_\oplus$ are considered likely to capture nebular gas during their formation and evolve into gas- or ice-giants. And planets between 1 and 10 $M_\oplus$, so-called "super-Earths", are expected to have plate-tectonic behavior different from Earth's.[8]

An accurate estimate of a planet's orbit is likewise critical to the issue of habitability. The stellar energy flux incident on a planet drives its atmospheric chemistry to a disequilibrium state. But the degree of disequilibrium can be profoundly increased by the presence of life on a planet's surface. As a prime example, the simultaneous presence of $O_2$ and reduced gases such as $CH_4$ or $N_2O$ in a planet's atmosphere is considered the best available remote evidence for Earth-like life[9]. Knowledge of a planet's orbit, including its eccentricity, can also provide an estimate of internal tidal heating, which might help drive plate tectonics.

The masses and orbits of planets around a star are critical to knowing whether the system is dynamically stable, and they provide clues about the evolutionary history of the system. As with stars, knowledge of the masses and orbits of planets within a system can be used to judge the validity of theoretical planet formation scenarios.

*Spectrum of Solar-System Twin at 10 pc*

If a nearby Solar System twin were to be found, it would have the broad spectral features shown in Fig. 1. For the visible and infrared wavelength regions, this figure gives the number of photons per second and per square meter, in a 10% bandwidth, that we would receive from twins of the Sun, Earth, and Jupiter. As the exozodiacal light is an extended source, it is shown for two cases: low angular resolution with a 1-m diameter telescope, and high angular resolution with an 8-m diameter telescope. In both cases the curves represent the total of local plus exozodiacal light in a spatial resolution element of the telescope. The smooth curves are black-body approximations. The spectral features for Earth and Jupiter are calculated from models which are based on actual spectra.

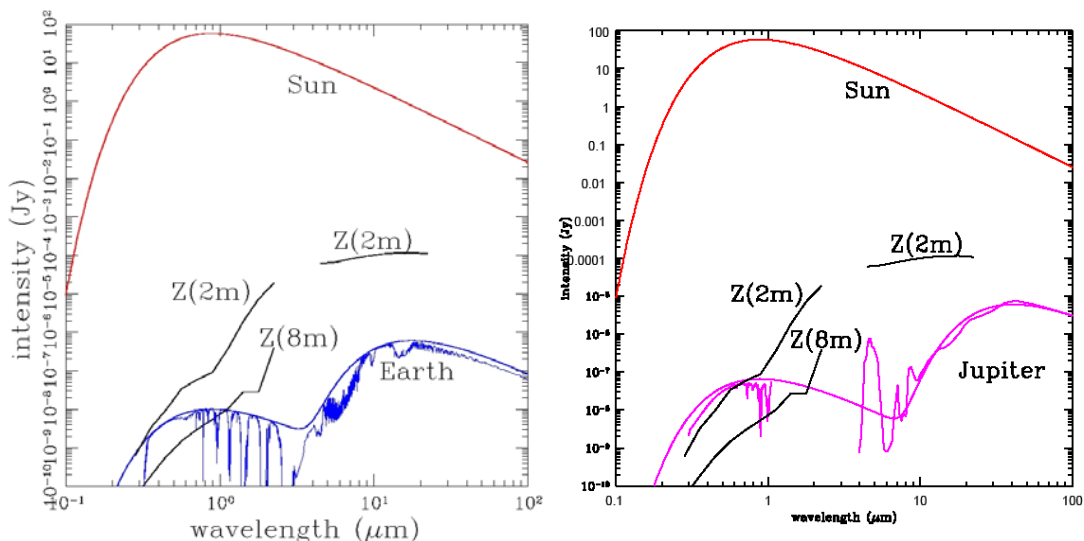

*Fig. 1 Solar-System twin spectra at 10 pc, schematic. (Left) Earth-twin spectrum with continuum and cloud-free absorption features. The exozodiacal light (from M. Kuchner) is sketched for small and large telescopes for reflected light, and an interferometer with 2-m collectors in the infrared.. (Right) Jupiter-twin spectrum and zodiacal light, same telescopes and interferometer, with visible (Karkoschka) and thermal infrared (Fortney) spectral features.*

These spectra are samples of what we would expect to see with direct imaging in the visible or infrared. (We should be careful here, because of course it is sometimes what we do not expect— hot Jupiters, for example—that turns out to be most interesting!) The Jupiter-twin is an easier target than the Earth-twin, owing to its greater angular separation from the star and to its higher flux levels. We note in passing that the contrast ratio (planet/star) of the Earth-twin is about $10^{-10}$ in the visible and $10^{-7}$ in the infrared, and that the nuisance of zodiacal light is minimized with a larger telescope, which also means that random and systematic noise from this source is reduced. In addition to its visualization value, Fig. 1 also allows us to make simple estimates of the expected signal-to-noise ratio for the measurement of continuum and spectral features, assuming that the limits are set by photon counting statistics. In the following sections we discuss the top-level science that could be obtained from direct-image spectra of nearby Earth-, Jupiter-, and zodi-twins.

*Earth-twin spectrum*
The Earth's spectrum consists of two different parts. From the left side of the figure out to about 4 µm, the flux from Earth is proportional to that from the Sun because, at these wavelengths, one is seeing the Earth in reflected sunlight. Longward of 4 µm, one sees the Earth's own thermal-infrared radiation. Earth's spectrum is crudely similar to that of a blackbody with a temperature of ~255 K. But there are absorption features, caused by the presence of various gases in Earth's atmosphere that are visible even at low spectral resolution. We ask the question: What could we see if we looked at the modern Earth from a great distance and what could we learn about the presence of life?

Many spectra of Earth have been taken from low-Earth orbit, but single-pixel, whole-Earth spectra are difficult to obtain. One way in which this has been done in the visible and near infrared is by using *Earthshine* data from the Moon.[10] The Moon's dark side is faintly illuminated by sunlight that has been reflected from Earth. So, if we observe the lunar dark side (from the ground) and then divide by the spectrum of the day side, we obtain a spectrum of Earth. This spectrum is an average over the entire sun-lit Earth as seen from the Moon. In the infrared several space missions have looked back at the Earth and taken its spectrum[23]. Examples of such Earthshine and infrared space instrument spectra are shown in Fig. 2.

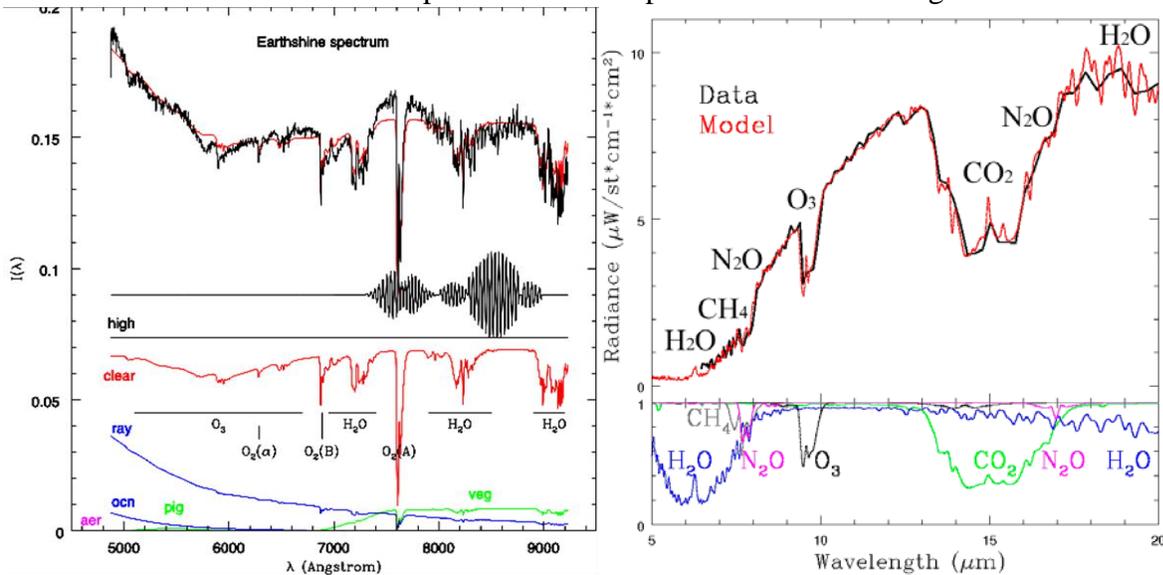

*Fig. 2 (left) Visible spectrum of the Earth, from "Earthshine" data.[10] and thermal infrared spectrum of Earth from the TESS instrument.[22, 23] See text for details.*

Although it appears dim to the naked eye, Earthshine actually contains lots of photons, and so one can easily subdivide it to create a medium-resolution ($R = 600$) spectrum (Fig. 2). In the figure, the wiggly black curve at the top represents the data. In the visible, the smooth curve running through the data is the sum of the "clear sky" and other curves drawn below it. In the thermal infrared the modeled contributing gases are likewise shown below the data. From this, one can see that three different gases can be detected in Earth's atmosphere in the visible: $O_2$, $O_3$, and $H_2O$ and three in the infrared with low resolution (25): $O_3$, $CO_2$, $H_2O$ and potentially two more with higher resolution (150): $N_2O$ and $CH_4$. $O_2$ itself has three different absorption bands that are visible at this spectral resolution. The strongest of these is the $O_2$ 'A' band at 0.76 µm. This band can be observed at relatively low spectral resolution (R=70) and was identified almost 30 years ago as a possible indicator of life on exoplanets[11]. The $O_2$ 'B' band at 0.69 µm is also easy to pick out, as are the three $H_2O$ absorption bands at 0.72, 0.82, and 0.94 µm.

Ozone, $O_3$, has one broad band in the visible (the Chappuis band) that extends from 0.5-0.7 µm. $O_3$ absorbs more strongly at shorter UV wavelengths, particularly in the Hartley band (0.2-0.3 µm), but one cannot see that wavelength region in this particular spectrum. Rayleigh scattering makes the planet appear blue, and its strength gives the total column density of molecules. From this, the planet's surface pressure can potentially be estimated, assuming that the surface gravity is known from mass and radius, although this result could be sensitive to clouds and to wavelength-dependent variations in surface albedo.

Another possible bioindicator may be present in the Earthshine spectrum in Fig. 2. In the region 0.70-0.75 µm one can see a modest increase in the amount of reflected light towards longer wavelengths. This increase may reflect the *red edge* of the chlorophyll molecule and leaf structure. The leaves of land plants reflect sunlight much more efficiently longward of this edge than they do in the visible.[12,13] Marine plants and algae do this as well, although the effect is muted by overlying water. This red edge is easy to pick out if one looks directly at a leaf, or down from space at a patch of densely vegetated land.[14] It is less visible in a disk-integrated spectrum like shown in Fig. 2 because most of the Earthshine light was reflected originally from the Pacific Ocean, where the vegetation signal is largely absent. On Earth about 70% of the surface is water, diluting the signal of surface features. The vegetation signal might also appear at a different wavelength, or not at all, in the spectra of alien vegetation.[13] On Earth widespread vegetation has been present since about 450 million years ago. Hence the red edge is a potential, but not necessarily robust, bioindicator.

Even more interesting is the strong band of $O_3$ centered at 9.6 µm. This band is clearly visible, even though ozone is only a trace gas in Earth's atmosphere. The reason for this is two-fold: First, the band itself is intrinsically strong, like the 15-µm band of $CO_2$. And, second, most of Earth's ozone is located up in the relatively cool stratosphere, making it easy to observe in absorption above the warmer troposphere and surface. The information obtained from observing $O_3$ is partially redundant with that gained from observing in the visible, as $O_3$ is produced photochemically from $O_2$. However, because of nonlinearities in ozone photochemistry, $O_3$ can be detected even if only small amounts of $O_2$ are present.[16,17] The thermal infrared $CO_2$ band at 15 µm is also clearly visible in the Earth's spectrum, even though the $CO_2$ concentration is relatively low, ~380 ppmv (parts per million by volume). The absorption at short wavelengths (< 8 µm) is caused by $H_2O$, as is the absorption at long wavelengths (> 17 µm). So, as in the visible, it should be possible to determine whether a planet has water vapor in its atmosphere.

Nearly all of Earth's $O_2$ comes from photosynthesis, which is carried out by plants, algae, and cyanobacteria. Predicted atmospheric $O_2$ concentrations prior to the origin of photosynthesis are too low to detect spectroscopically.[15] Hence, the observation of $O_2$ in an extrasolar planet's atmosphere would be, under most circumstances, strong evidence for the existence of life on that planet. Ultimately, as noted earlier, we will want to look for the simultaneous presence of both reduced and oxidized biogenic gases[9], but this will require telescopes bigger than we envision for the near future. On the early Earth $O_2$ did not become abundant until about 2.4 Gyr ago, based on isotopic and geologic evidence.[18] $O_3$ should have been scarce, as well, because it is formed from $O_2$. But life has probably existed on Earth since at least 3.5 Gyr ago, and possibly earlier. Would we be able to tell this if we observed an early Earth-type planet in the visible or infrared? Models can give us some ideas[22]. The most obvious gas to look for is methane, $CH_4$. Methane was probably relatively scarce prior to the origin of life, as the prebiotic atmosphere is thought to have been dominated by $N_2$ and $CO_2$.[19] Atmospheric $CH_4$ concentrations probably increased dramatically once methanogenic bacteria evolved and began generating $CH_4$ biologically. Today's $CH_4$ concentration is relatively low, 1.7 ppmv, but prior to the rise of $O_2$ the methane lifetime would have been longer and its concentration could have been 1000 ppmv or more.[20] This amount of $CH_4$ could be measured in the visible (several bands) as well as in the thermal-infrared (7.7-μm). $CH_4$-rich atmospheres can also generate Titan-like organic hazes, which might be observable spectroscopically. So, if we find a nearby planet like the post-biotic Archean Earth, we should be able to identify it as such from its spectrum. Whether this would be interpreted as a sign of life is not clear, but it would certainly generate a huge amount of debate.

*Jupiter-twin spectrum*
The spectrum of a Jupiter twin is easily distinguishable from that of an Earth. For example, Jupiter has strong $CH_4$ bands in the red visible spectrum, and no $CO_2$ or $O_3$ features at any wavelength. At gigayear ages Jupiter's spectrum will be roughly independent of its mass, but will vary significantly with distance from its star, owing to the formation of water, methane, or ammonia clouds at different altitudes, and depending on the temperature structure of the atmosphere and the relative abundance of metals. Giants somewhat warmer than Jupiter would be quite bright, with spectra controlled by high, thick water clouds punctuated by the strongest methane bands. Even warmer giants (younger, more massive, or closer to their primary star) may lack clouds altogether, leading to very blue, Rayleigh-dominated spectra and little scattered red flux. At solar system ages the flux shortward of about 3 to 4 μm is entirely scattered incident light. The giants are particularly bright in thermal emission at 5 μm (where there is a hole in the water opacity, allowing flux from deep, warm layers to escape), and all flux at longer wavelengths is thermal emission. Goals for giant planet characterization include estimating the planet's mass and measuring atmospheric composition, looking in particular to see if it is enhanced in heavy elements. All the solar system giants are enhanced in heavy elements over solar abundance, and the pattern of enrichment is commonly interpreted as signature of the giant planet formation process. Specifically, disk instability models predict metallicities close to that of the host star, whereas the core accretion mechanism leads to significantly enhanced metallicity. Discerning the enrichment pattern in other planetary systems will provide important new insights into planet formation.

*Zodi-twin spectrum*

In the background, literally, is the analog of our Solar System's zodiacal cloud of dust and gas. This exozodiacal cloud is both a help and a hindrance to exoplanet observations, as it gives clues about the existence, architecture, and evolution of the system, but at the same time potentially obscures the planets from visible and infrared observations. The spectrum contains information on the composition and grain size distribution of the dust cloud, which in turn helps to illuminate the nature of the larger dust-producing bodies. Spectra of dust clouds will also aid in revealing the density distribution of the dust cloud and possibly any unresolved structure, such as circumstellar rings. Spatial resolution of a system's zodiacal dust cloud may reveal azimuthal and radial structure. These structures can inform us of the distribution of bodies creating the dust, such as a belt of bodies analogous to our own asteroid belt. Terrestrial mass planets can also sculpt clumpy circumstellar ring structures in the dust, which we can use to infer the presence of unseen planets and to characterize their mass and orbital parameters. Exo-Earths and even planets as small as a few times Mars' mass are detectable through the resonant structures in exozodiacal clouds at a distance from the host star of about 10 AU.[21] All in all, exozodiacal clouds are sufficiently important that they should be studied intensively from the ground and from space, both for their own intrinsic scientific merit and to help us understand what is needed to directly image planets.

*Summary*

The scientific gain expected from the detection and characterization of nearby exoplanets is enormous. The further gain if we see any sign of life is beyond estimation. It will affect our entire perception of our place in the universe, thereby extending the Copernican revolution. The science to be extracted from exoplanet spectroscopy depends on light-collecting ability and angular resolution, and so instrumentation is inextricably intertwined with science success. Space-based astrometry and direct imaging missions currently under study may already have the capability to do this, and we hope that they will be pursued vigorously.